\documentclass[journal=ancac3,layout=twocolumn,amsmath,amssymb]{achemso}

\usepackage{achemso}

\usepackage[T1]{fontenc}
\usepackage{graphicx}

\usepackage{amsmath}
\usepackage{mhchem}
\usepackage{times}
\usepackage{multirow}
\usepackage{pdfpages}

\newcommand{\rb}{\mathbf{r}}

\newcommand{\kb}{\mathbf{k}}

\newcommand{\ub}{\mathbf{u}}

\newcommand{\eb}{\mathbf{e}}

\newcommand{\etal}{\textit{et al}.}
\newcommand{\ie}{\textit{i}.\textit{e}.}
\newcommand{\eg}{\textit{e}.\textit{g}.}
\newcommand{\icm}{cm$^{-1}$}

\newcommand{\overlap}[2]{\langle#1|#2\rangle}

\usepackage{xcolor}
\usepackage{cuted}

\title{Raman Spectra of 2D Titanium Carbide MXene from Machine-Learning Force Field Molecular Dynamics}
\author{Ethan Berger}\email{ethan.berger@oulu.fi}
\affiliation{Microelectronics Research Unit, Faculty of Information Technology and Electrical Engineering, University of Oulu, P.O. Box 4500, Oulu, FIN-90014, Finland}

\author{Zhong-Peng Lv}
\affiliation{Department of Applied Physics, Aalto University, Aalto, FIN-00076, Finland}

\author{Hannu-Pekka Komsa}\email{hannu-pekka.komsa@oulu.fi}
\affiliation{Microelectronics Research Unit, Faculty of Information Technology and Electrical Engineering, University of Oulu, P.O. Box 4500, Oulu, FIN-90014, Finland}

\begin{document}
\begin{strip}
\begin{abstract}
MXenes represent one of the largest class of 2D materials with promising applications in many fields and their properties tunable by the surface group composition. Raman spectroscopy is expected to yield rich information about the surface composition, but the interpretation of measured spectra has proven challenging. The interpretation is usually done via comparison to simulated spectra, but there are large discrepancies between the experimental and earlier simulated spectra.
In this work, we develop a computational approach to simulate Raman spectra of complex materials that combines machine-learning force-field molecular dynamics and reconstruction of Raman tensors via projection to pristine system modes. The approach can account for the effects of finite temperature, mixed surfaces, and disorder.
We apply our approach to simulate Raman spectra of titanium carbide MXene and show that all these effects must be included in order to properly reproduce the experimental spectra, in particular the broad features. We discuss the origin of the peaks and how they evolve with surface composition, which can then be used to interpret experimental results.
\end{abstract}
\end{strip}

\newpage

\

\newpage

\

\newpage

\begin{figure} 
    \centering
    \includegraphics[width=8.25cm]{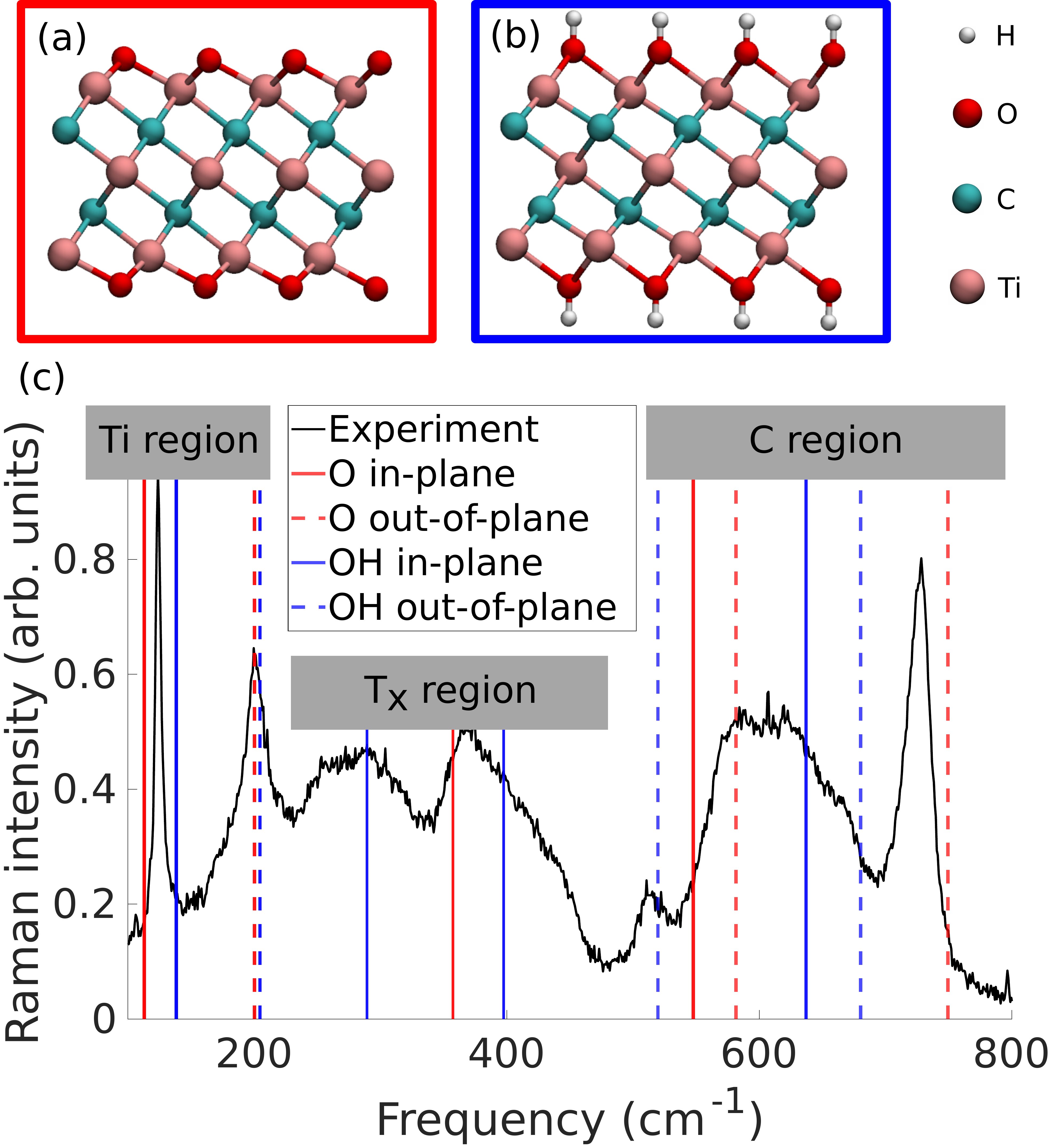}
    \caption{(a)-(b) Balls-and-sticks representation of pure \ce{Ti3C2O2} and \ce{Ti3C2(OH)2}, respectively. Titanium is in pink, carbon in blue, oxygen in red and hydrogen in white. (c) Experimental Raman spectrum of \ce{Ti3C2T_{\rm{x}}} compared with the frequencies of the Raman active modes calculated using DFT. The red and blue lines correspond to calculations of pure \ce{Ti3C2O2} and pure \ce{Ti3C2(OH)2}, respectively. Full lines represent the in-plane modes while the dashed ones represent the out-of-plane modes. }
    \label{fig:Fig1}
\end{figure}

Two-dimensional (2D) materials have attracted a lot of attention from the scientific community in the last two decades. Among them, MXenes represent one of the largest class of 2D materials with more than 40 synthesized compositions and many more still left to discover \cite{Naguib_2021}. Due to their high electrical conductivity, MXenes show promise for applications in various fields, including energy storage, gas sensors, and electromagnetic interference shielding \cite{Anasori_2017,Lee_2019, Iqbal_2020}. MXenes take the form \ce{M_{\rm{n+1}}X_{\rm{n}}T_{\rm{x}}} where M is an early transition metal (\eg{} titanium), X can be carbon or nitrogen, T are the surface terminations which depend on the synthesis methods, and $\rm{n}=1$--4 defines the thickness of the layer. The first synthesized and the most studied MXene \ce{Ti3C2T_{\rm{x}}} is obtained by selective etching of the Al layer from its precursor bulk phase \ce{Ti3AlC2} \cite{Naguib_2011}. Due to this process, the surfaces are passivated by functional groups from the etching solution, with the most common ones being -O, -OH and -F. Balls-and-sticks representation of pure -O and -OH surfaces of \ce{Ti3C2} are represented in Fig.\ \ref{fig:Fig1}(a,b), respectively. Under the common wet-etching synthesis conditions, the surface is never purely O or OH but contains a randomly arranged mixture of these functional groups \cite{Mashtalir13_MCP,Shi14_PRL,Wang16_CM,Ibragimova19_ACSNano,Ibragimova21_JPCL}
Unfortunately, it has turned out to be challenging to not only control the ratio of these terminations, but also to quantify them.

One commonly used method to study the composition of 2D materials is Raman spectroscopy, which is a non-destructive method for characterizing the vibrational properties of molecules or solids. Raman spectra of \ce{Ti3C2T_{\rm{x}}} has been measured and carefully studied experimentally \cite{Sarycheva_2020,Lioi_2019,Sarycheva22_ACSNano}. To explain the experimental observations, Raman spectra have been calculated using density functional theory (DFT) for pure surfaces \cite{Hu_2015,Hu_2018}, but the comparison is not straightforward. To illustrate these problems,
in Fig. \ref{fig:Fig1}(c) we reproduce typical experimental Raman spectrum and the frequencies of Raman active modes of pure surfaces (-O and -OH) from DFT (description of the synthesis and Raman measurement for the experimental results are available in the Supporting Information). For the modes around 120 cm$^{-1}$, 200 cm$^{-1}$, and 700 cm$^{-1}$, the experimental frequencies appear to be close to, or in between, those of the calculations for pure surfaces, and thus these peaks can be quite safely assigned to the calculated modes. 
More importantly, other peaks around 300-400 cm$^{-1}$ and 600 cm$^{-1}$ both show extremely broad features which cannot be explained by static calculations of pure surfaces. There are few possible origins for this discrepancy.
First, static calculations give no information about the width of the Raman peaks. 
Calculation of anharmonic interatomic force constants \cite{Lindsay2013} or molecular dynamics (MD) simulations are usually necessary to account for temperature effects and obtain realistic line widths. In the latter, Raman intensities can be obtained from the Fourier transform of the polarizability $\chi(t)$ autocorrelation function \cite{Thomas13_PCCP}, but requires tens of thousands of polarizability calculations.
It has successfully been used to compute Raman spectra of solid ice \cite{Putrino02_PRL}, liquid water \cite{Wan13_JCTC}, and molecules \cite{Luber14_JCP}, but such calculations are very expensive and usually restricted to a small number of atoms. Secondly, wide peaks of the MXene Raman spectra might come from the structural disorder, \ie{}, the mixed surfaces. In order to simulate the different surface terminations, it is necessary to use large supercell, making the calculation of polarizabilities at every time step of an MD run computationally intractable. 

With the recent developments in machine-learning force fields (MLFF) it has now become possible to carry out MD simulations for long times and large number of atoms efficiently, yet still nearly matching the accuracy of first-principles calculations.
\cite{Kocer_2022,Behler_2007,Bartok_2010,Fan_2021,Jinnouchi_2019_1,Jinnouchi_2019_2}. Moreover, by using MLFF, it becomes possible to run MD over a large configurational space, such as different compositions and distributions of surface terminations in MXenes. Unfortunately, MLFF usually cannot be used to evaluate polarizabilities as they only model energies and forces. To compute the polarizabilities of these large supercells, Hashemi \etal{} recently developed a computational scheme (denoted RGDOS), which allows to obtain the Raman tensors of a supercell from a projection onto those of the unit cell \cite{Hashemi19_PRM}. The scheme has already been successfully applied to, \eg{}, transition metal dichalcogenides alloys (\ce{Mo_xW_{1-x}S2} and \ce{ZrS_xSe_{1-x}}) \cite{Hashemi19_PRM,Oliver20_JMCC}, defects in \ce{MoS2} \cite{Kou_2020}, and SnS multilayer films \cite{Sutter21_NT}. Adapting RGDOS for evaluating time-dependent polarizability $\chi(t)$ and combining it with MLFF MD trajectories
would lead to a highly efficient scheme to obtain the Raman spectra at finite temperature and in large supercells. 

In this work, we present a scheme for combining MLFF trajectories with RGDOS (or DFT calculated Raman tensors) to evaluate finite-temperature Raman spectra of complex systems. The method leads to highly efficient MD runs and calculations of the polarizabilities, allowing us to easily probe large configurational space. We apply the method to 2D titanium carbide MXene to study the effect of heterogeneous surfaces on the Raman spectra and attempt to give a better description of the experimental results. We first investigate the effects of temperature and mixed surface terminations on the Raman spectrum, and finally study the effect of additional disorder by considering the vibrational modes outside $\Gamma$-point. 
We discuss the origin of the peaks and broad features and how they evolve with the surface composition.


When using MD, the Raman spectrum is obtained as the Fourier transform of the autocorrelation function of polarizability $\chi$: \cite{Thomas13_PCCP,Medders_2015}
\begin{equation}
I(\omega) = \int \langle\chi(\tau) \chi(t+\tau)\rangle_\tau \ e^{-i\omega t} dt
\label{equ:RamanMD}
\end{equation}
where $\langle x(\tau) x(t+\tau)\rangle_\tau$ denotes the autocorrelation function. The challenge therefore is how to obtain the polarizability of a large supercell for every configuration visited during MD at a reasonable computational cost. In RGDOS \cite{Hashemi19_PRM}, the Raman tensors of a large supercell are obtained by first projecting its vibrational modes onto those of the pristine unit cell and then combining the Raman tensors of the unit cell weighted with these projections. This means that only the Raman tensors of the unit cell have to be calculated, resulting in an immense reduction of computational effort. The scheme has already been successfully applied to study the effect of alloys, defects, and film thickness on the Raman spectra, all involving large supercells \cite{Hashemi19_PRM,Kou_2020,Oliver20_JMCC,Sutter21_NT}. Since this method does not rely on the electronic structure of the supercell, it is perfectly suited to be used with empirical or MLFF MD simulations. Moreover, when combined with MD simulations, such method would also account for temperature effects and anharmonicity.

In this work, we adopt an approach conceptually similar to the RGDOS. 
We start by writing the polarizability as a Taylor expansion around the relaxed positions $\rb_0$, which leads to the following equation where $\chi_0$ is the polarizability at $\rb_0$ and $\ub$ represent the atomic displacements
\begin{equation}
    \chi(\ub) = \chi_0 + \frac{\partial \chi}{\partial \ub} |\ub|+ O(|\ub|^2).
    \label{equ:chi(u)}
\end{equation}
One can notice the resemblance between the partial derivative $\frac{\partial \chi}{\partial \ub}$ and the definition of Raman tensors $R_m=\frac{\partial \chi}{\partial \eb_m^*}$ (in the Placzek approximation\cite{placzek1934,Long2002-qo}), where $\eb_m^*$ are the mass-scaled eigenvectors of the unit cell.
This suggests that we could use the set of eigenvectors $\{\eb_m^*\}$ as a basis for the expansion, with Raman tensors giving the expansion coefficients and the "distance" given by the projection of $\ub$ to $\eb_m^*$, written out as
\begin{equation}
    \frac{\partial\chi}{\partial\ub}|\ub| = \sum_m R_m P_{m}(\ub).
    \label{equ:RGDOS}
\end{equation}
where $P_m(\ub)$ is the projection of $\ub$ onto $\eb_m^*$.
We first note, that since $\eb_m^*$ do not form an orthogonal basis, the projection coefficients $P_m(\ub)$ have to be found by solving the following 
system of linear equations
\begin{equation}
    \sum_i\overlap{\eb_m^*}{\eb_i^*}P_i(\ub) = \overlap{\eb_m^*}{\ub}.
    \label{equ:MDproj}
\end{equation}
Moreover, since only $\Gamma$-point modes contribute to first-order Raman scattering, the projections are here done from the supercell displacements $\ub$ to the unit cell eigenmodes at $\Gamma$-point $\eb_m^*$. 

The final equation for the time-dependent polarizability is then
\begin{equation}
    \chi(t) = \chi(\ub(t)) = \chi_0 + \sum_{m} R_m P_m(\ub(t)).
    \label{equ:Pol_RGDOS}
\end{equation}
When the eigenmodes and the Raman tensors of the unit cell and the MD trajectories are known, $\chi(t)$ can easily be obtained from equations \ref{equ:MDproj} and \ref{equ:Pol_RGDOS}. This method is therefore highly efficient and compatible with MLFF MD.
We also note that in equation \ref{equ:Pol_RGDOS}, the second order term could be added, essentially corresponding to the second-order Raman scattering. In this case, the sum should also be carried over different points of the first Brillouin zone (BZ) to account for the modes outside of Gamma-point. In the case of MXenes, there is no indication of dominant second order contribution and thus we have decided not to include the second order term.


\begin{figure*} 
    \centering
    \includegraphics[width=16.5cm]{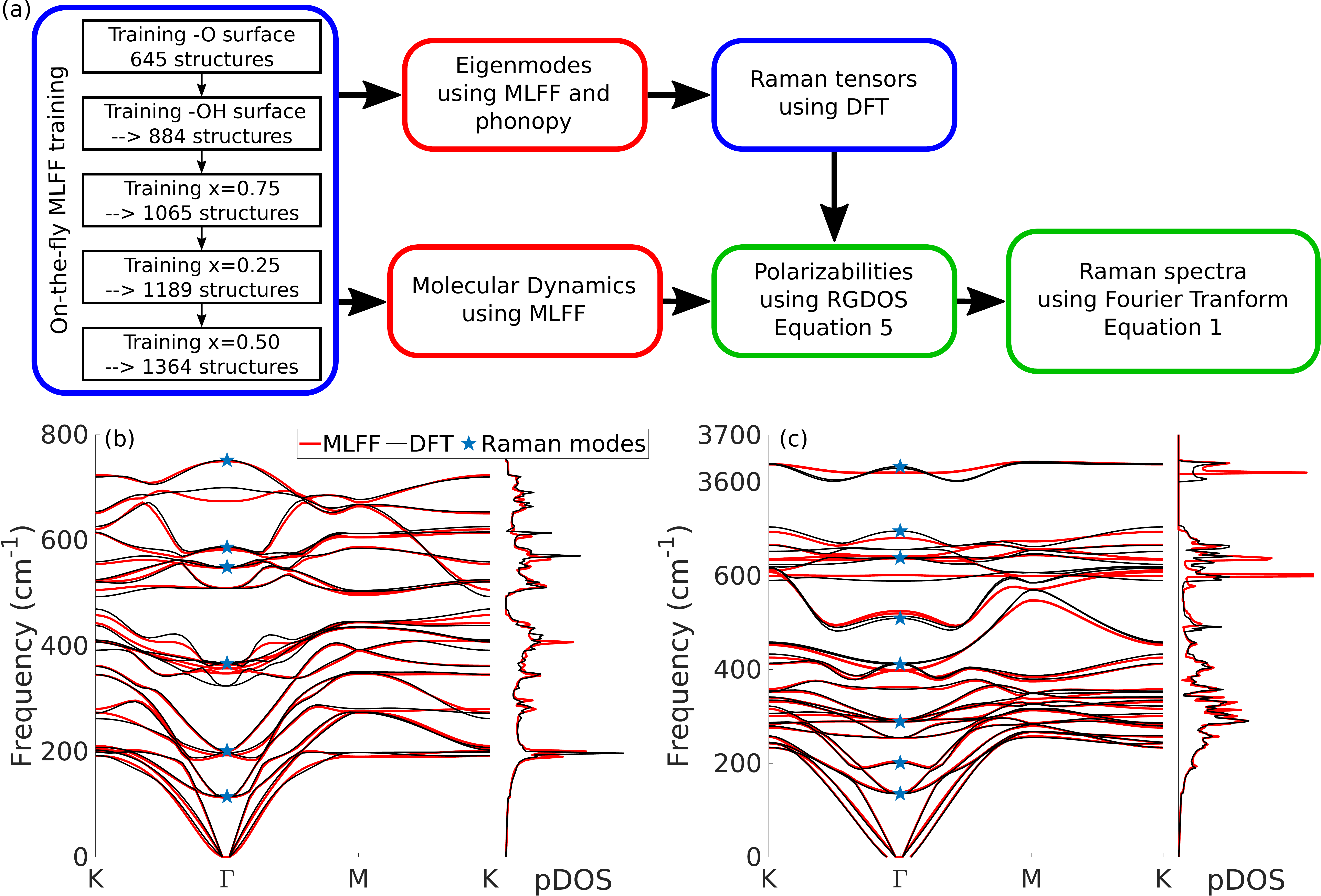}
    \caption{(a) Schematic workflow of the computational approach. Blue boxes represent steps using DFT calculations, red using MLFF, and green are post-processing steps. (b)-(c) Phonon dispersion curves and density of states (pDOS) of \ce{Ti3C2O2} and \ce{Ti3C2(OH)2}, respectively. Results from MLFF (red lines) are compared with those from DFT (black lines). Blue stars indicate Raman-active modes at $\Gamma$-point. }
    \label{fig:Fig2}
\end{figure*}

All calculations are done using the Vienna \textit{ab initio} simulation package (VASP) \cite{Kresse_1996_1,Kresse_1996_2}. We adopt the Perdew-Burke-Ernzerhof exchange-correlation functional for solids (PBEsol) \cite{Perdew_2008} and set the plane-wave cutoff to 550 eV. Phonon eigenmodes are obtained using the Phonopy software \cite{Togo_2015}. For determining the force constants, we used 4$\times$4 supercells with the Brillouin zone sampled using a 4$\times$4 k-point mesh. Once the eigenvectors are known, the Raman tensors were obtained from a centered finite difference scheme. 
Since MXenes are metallic, the polarizabilities have to be evaluated at finite excitation energy $\omega$, i.e., using $\chi(\omega)$ instead of $\omega \to 0$ limit normally used in nonresonant Raman. Although Placzek approximation is derived under non-resonant conditions, it has been found to work well also in resonant conditions \cite{Walter20_JCTC}.
Here, for the projections we used the eigenvectors and Raman tensors of -OH unit cell, the latter evaluated at $\omega=516$ nm. More information regarding resonant Raman and the choice of Raman tensors are available in the Supporting Information (see in particular Fig.\ S1). The frequency-dependent polarizability (or dielectric function) is computed using a summation over empty bands \cite{Gajdos_2006}. To guarantee accurate evaluation of small changes in $\chi(\omega)$, a denser k-point mesh of 48$\times$48 (in the unit cell) was adopted and the number of orbitals was increased to 120 (roughly 4 times the recommended amount). For phonon modes and Raman tensors calculations, the electronic structure was relaxed with a precision of 10$^{-7}$ eV. 

The machine learning force field (MLFF) was trained using on-the-fly machine learning, as recently implemented in VASP \cite{Jinnouchi_2019_1,Jinnouchi_2019_2}. This method has already been successfully applied to different materials including different phases of zirconium \cite{Liu_2021} and hybrid perovskites \ce{MAPbX3} \cite{Bokdam_2021,Lahnsteiner_2022}. 
Fig.\ \ref{fig:Fig2}(a) shows a schematic of the workflow and how MLFF, MD, DFT and RGDOS are all combined to obtain the Raman spectrum.
In the training, we again used 4$\times$4 supercells with a 4$\times$4 k-points mesh, but the electronic structure convergence criterion was loosened to 10$^{-4}$ eV. The model was first trained for pure surfaces before being trained for mixed -O and -OH surfaces as well.
F-terminated surfaces were not included in order to reduce the complexity of the MLFF model and since they are expected to behave fairly similar to OH-groups \cite{Xie13_PRB,Khazaei17_JMCC,Ibragimova21_JPCL}. The vibrational frequencies are generally bit lower than for OH-surface \cite{Hu16_PCCP} due to the slightly larger mass of F. 
Each step consisted of MD simulation in the NVT ensemble at 300 K for 20 hours (40000--70000 steps), and the MLFF configurations after each step carried on to the next step. In total, the training set contains 1364 configurations. 
The root mean square error (RMSE) between DFT and MLFF forces is 45.8 meV/\AA.
To benchmark the model, we compared the phonon dispersion curves from DFT and the MLFF model. Fig. \ref{fig:Fig2}(b) and (c) show the phonon bands and density of states (pDOS) for pure \ce{Ti3C2O2} and \ce{Ti3C2(OH)2}, respectively. Results are overall in good agreement for the dispersion curves as well as for the pDOS. 
Fortunately, the few modes where errors at $\Gamma$-point are somewhat larger, such as the 350 and 700 \icm{} modes in \ce{Ti3C2O2}, happen to be Raman-inactive modes.
We also note, that since MLFF is fitted into finite-temperature trajectories, the 0 K force constants used in the phonon dispersion curves may differ slightly from the DFT results (the effect of temperature is discussed below).
Those benchmarks indicate that the model is sufficiently trained and that we can use it for reliable MD production runs.

Molecular dynamics were performed using the MLFF model in the NVT ensemble using a Nosé-Hoover thermostat \cite{Nose_1984,Nose_1991} for 8x8 supercells. The time step was set to 1 fs and 10$^5$ steps were performed, resulting in runs of 100 ps. To reduce noise in the final spectra, five runs with different random distributions of the surface terminations are performed for each concentration. Using different distribution of the surface termination does not impact the main features of the resulting Raman spectra, see Fig.\ S2 in the Supporting Informations for more details.


\begin{figure} 
    \centering
    \includegraphics[width=8.25cm]{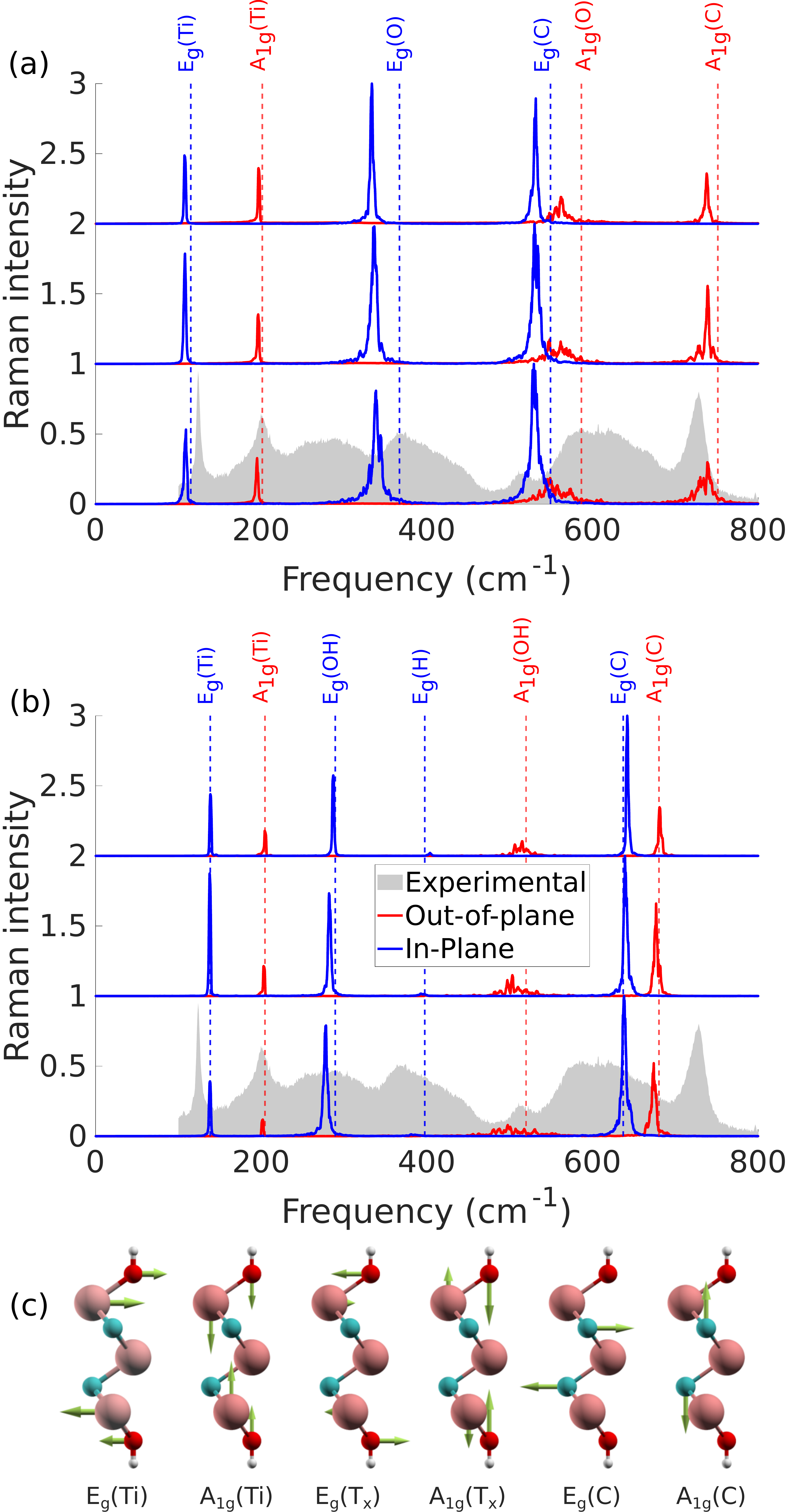}
    \caption{(a)-(b) Raman spectra at different temperature for \ce{Ti3C2O2} and \ce{Ti3C2(OH)2}, respectively. Blue lines represent the in-plane modes while red ones are the out-of-plane modes. The grey areas show experimental results and the dashed vertical lines show the frequencies from phonon calculations using MLFF. (c) Eigenmodes of Raman active modes.}
    \label{fig:Fig3}
\end{figure}

We first investigate the effect of temperature on the width of the peaks. Fig.\ \ref{fig:Fig3}(a) and (b) show the Raman spectra of \ce{Ti3C2O2} and \ce{Ti3C2(OH)2} at different temperatures, respectively. For a better understanding of the different vibrations, Fig.\ \ref{fig:Fig3}(c) shows a representation of the Raman-active modes. The peak frequencies compare well to the DFT results presented in Fig.\ \ref{fig:Fig1}(c) for both systems, showing that our approach leads to reasonable spectra. Overall, the frequencies do not shift markedly with temperature, but there is now a realistic description of the peaks widths. For some peaks, such as the two modes at 100 and 200 cm$^{-1}$ localized to titanium atoms, show weak dependence on temperature. While our results agree with experimental observations for the first, the width seems underestimated for the latter. For other peaks, the width increases with temperature. The A$_\text{1g}$ carbon mode at around 700 cm$^{-1}$ shows a fairly narrow peak at low temperature but it gets significantly wider with temperature, resulting in a better agreement with experiment at 300 K. Similar broadening with increasing temperature is observed for the in-plane modes of surface terminations and carbon at 300--400 and 500--600 cm$^{-1}$, respectively. Since surface terminations (-OH groups especially) should be strongly anharmonic, significant broadening could be expected for modes involving these atoms. In those cases however, the widths at 300 K are still underestimated and does not compare well to experiments. Overall, using finite temperature leads to a better comparison with experiments, but it is clear that something is still missing in order to correctly reproduce experimental observations.

\begin{figure} 
    \centering
    \includegraphics[width=8.25cm]{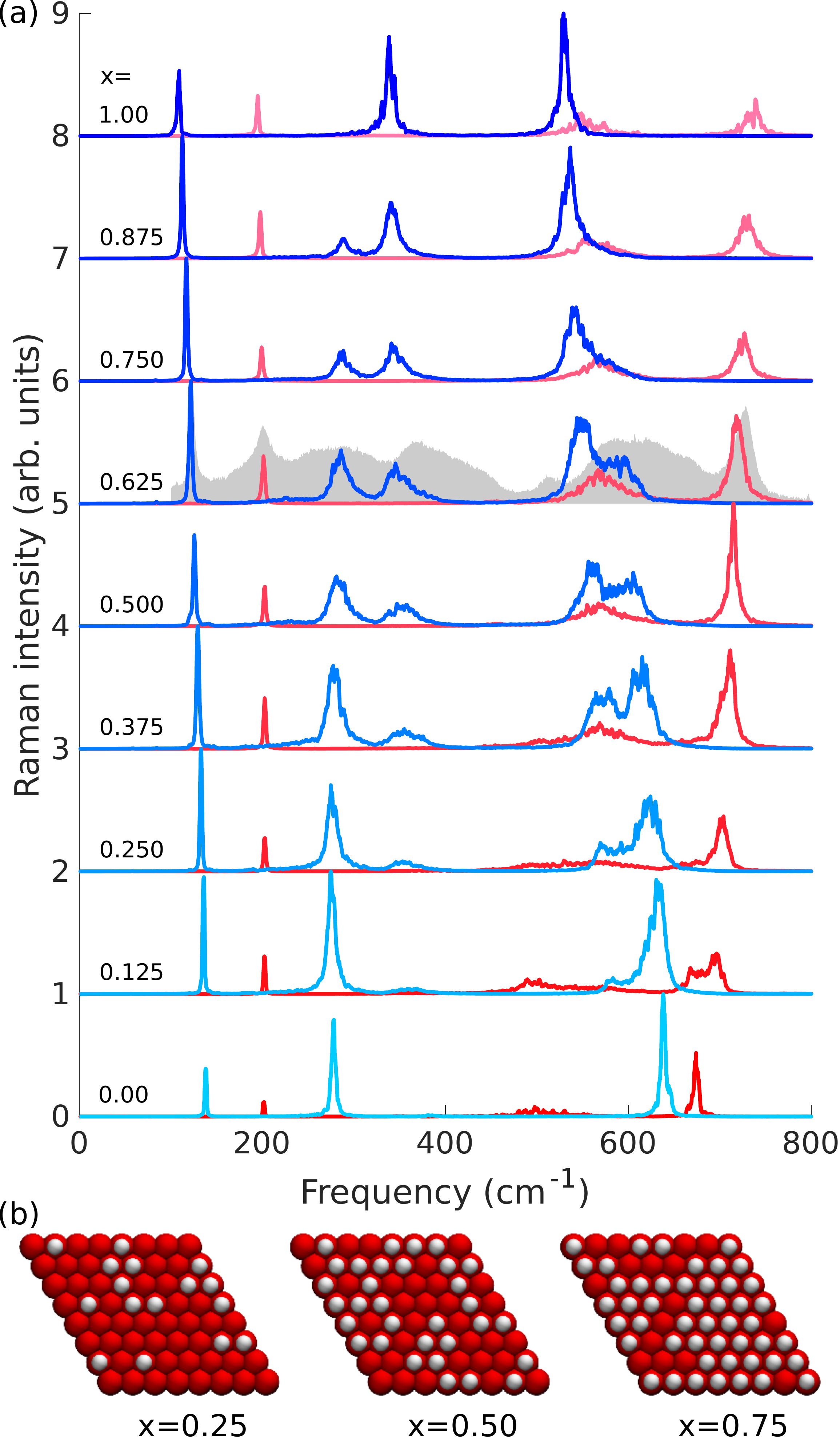}
    \caption{(a) Raman spectra of Ti$_3$C$_2$(O$_x$OH$_{1-x}$)$_2$ for different surface group compositions. Labels on the left show the concentration of oxygen at the surface, e.g. the top and bottom lines represent pure -O and -OH surfaces, respectively. In-plane modes are represented by the blues lines while out-of-plane modes are in red. Grey area show experimental observation for comparison. (b) Top view of the surface structure of Ti$_3$C$_2$(O$_x$OH$_{1-x}$)$_2$ for $x=0.25$, $0.5$ and $0.75$ (only shown one of the five configurations averaged over). Distribution of surface terminations is random and red balls show -O sites while white ones show -OH. }
    \label{fig:Fig4}
\end{figure}

We next look at the effect of mixed surfaces. Fig.\ \ref{fig:Fig4}(a) shows the Raman spectra for different concentrations of surface terminations and Fig.\ \ref{fig:Fig4}(b) illustrates the distribution of surface terminations for different concentrations. The top and bottom lines show results for pure \ce{Ti3C2O2} and \ce{Ti3C2(OH)2}, respectively. 
The titanium A$_\text{1g}$ mode at 200 \icm{} is largely unaffected by the composition changes.
The other two sharp peaks from the titanium E$_\text{g}$ mode at around 120 cm$^{-1}$ and the carbon A$_\text{1g}$ mode at around 700 cm$^{-1}$ both show a straightforward linear shift with O/OH ratio (a so-called one-mode behavior).
This suggests that these peaks could be used to find the surface termination composition from experimentally measured spectrum.
To this end, we extracted the frequencies by isolating the peaks and fitted them using a Lorentzian function (see Fig.\ S3 in the SI for details), and collected the results in Table \ref{tab:tab1} together with experimental values. In particular, we include the results from our experimental spectrum shown in Fig.\ \ref{fig:Fig1}(c) (and again in Fig.\ \ref{fig:Fig4}(a) below the x=0.625 spectrum) and those from Ref.\ \citenum{Sarycheva_2020}, where these peaks were found to shift depending on the sample preparation conditions.
First, if we use the Ti A$_\text{1g}$ mode as a probe for the accuracy of our calculations, we find that the calculated frequencies are 0--10 \icm{} lower than the experimental values. 
Second, by comparing the frequencies of the other two modes, best agreement with experiments is found for $x$ around 0.5--0.625. Even accounting for the inaccuracies in the calculated frequencies, our results strongly point to all these samples having surfaces with mixture of surface terminations. Interestingly, the composition range found from this comparison also agrees well with that predicted by first-principles calculations in Ref.\ \citenum{Ibragimova19_ACSNano}:
O$_{0.75}$OH$_{0.25}$ (x=0.75) at higher pH and
O$_{0.5}$OH$_{0.25}$F$_{0.25}$ at low pH
(corresponding to a total concentration of 0.5 for -OH and -F terminations).
For these two peaks, calculations also reproduce well the widths of the experimental peaks. 

In addition, there are the two in-plane modes at 300--400 \icm{} and 500--650 \icm{}, which show distinct two-mode behavior with changing composition.
For the former, it seems that the two broad features in the experimental spectrum arise from these two modes. For the latter, it is difficult to distinguish the two peaks from the experimental spectrum, but as this feature is very broad it is consistent with being comprised of two peaks as found in our calculations.
Also note that at concentrations close to x=0.5, the out-of-plane mode of carbon also seem to contribute at a similar frequency. 
Even though Raman spectra simulated using mixed surfaces agrees much better with experiments and would appear to contain all the main features, the agreement is clearly still far from perfect. In particular the width of some peaks, mainly those arising from the surface terminations and the carbon E$_\text{g}$ modes, are still greatly underestimated and by more than one might expect from instrumental broadening.

\begin{table*}
\begin{tabular}{cccccccccccc} \hline
                 & \multicolumn{9}{c}{Simulated}                                   & \multicolumn{2}{c}{Experiments}                     \\
      & 1.000 & 0.875 & 0.750 & 0.625 & 0.500 & 0.375 & 0.250 & 0.125 & 0.000 & Fig.\ \ref{fig:Fig1} & Ref.\ \citenum{Sarycheva_2020} \\ \hline
E$_\text{g}$(Ti) & 108.4 & 112.5 & 116.8 & 121.6 & 125.7 & 129.3 & 132.9 & 135.7 & 138.1 & 123.7 & 119.8 - 124.5 \\
A$_\text{1g}$(Ti) & 194.9 & 197.7 & 199.2 & 200.3 & 202.3 & 202.6 & 202.5 & 202.3 & 201.5 & 199.6 & 201.0 - 206.7 \\
A$_\text{1g}$(C) & 737.4 & 729.4 & 725.2 & 718.9 & 714.5 & 709.9 & 702.0 & 690.2 & 673.9 & 728.3 & 737.7 - 719.6 \\ \hline
\end{tabular}
\caption{\label{tab:tab1}Frequencies (in cm$^{-1}$) of the E$_\text{g}$(Ti), A$_\text{1g}$(Ti), and A$_\text{1g}$(C) modes for different ratios of O/OH surface terminations, and comparison to experimental values.}
\end{table*}

The peak broadening could arise from the interaction of surface terminations with water or other surface adsorbates, from the interaction with neighboring MXene layers, or from disorder arising, e.g., from point defects, grain boundaries, etc. 
We simulated the Raman spectra of multilayer MXene flakes, but this did not lead to marked broadening of these features as shown in Fig.\ S4(a) in the Supporting Information. Moreover, experimentally measured spectra of monolayer flakes, multilayer flakes, and colloidal dispersions all show largely similar broadening \cite{Lioi_2019,Sarycheva_2020,Sarycheva22_ACSNano}. We also investigated the presence of (small amount of) water on the surface, but also did not yield the desired effect (see Fig\ S4(b) in the SI).

One might expect that the wet chemistry and strong acids used in the synthesis would lead to a large number of defects, in addition to the defects inherited from the precursor MAX phases \cite{Zhang20_JMST,He19_CTC,Ibragimova22_CM}. Consequently there is also a large variety of possible defects: in addition to flake edges and grain boundaries \cite{Anasori_2017,Zhang20_JMST,Benitez16_AM}, the reported point defects include Ti vacancies and Ti adatoms \cite{Karlsson15_NL,Sang16_ACSNano}, C vacancies \cite{Mathis21_ACSNano}, and substitutional O in C site \cite{Tian22_CEJ}. 
Unfortunately, the dominant types of defects and their density is largely unknown, which makes simulating the role of defects on the Raman spectra challenging.

To circumvent this problem, we simulate the effect of disorder by considering the modes outside $\Gamma$-point.
In ideal crystals, first order Raman spectra only comes from modes at $\Gamma$-point because of the condition $\kb=0$ \cite{cardona1982}. However, in the presence of defects, the symmetry is lowered and the modes from the rest of the first Brillouin zone can contribute \cite{Ushioda_1974,Eckmann_2013}. 
In fact, this effect is already present in our calculations for randomly distributed surface terminations, but clearly the degree of disorder from surface terminations is insufficient.
To explicitly include the contributions from modes outside the $\Gamma$-point, we study here the phonon density of state using phonon-projected velocity-autocorrelation function (VACF)\cite{Wang90_PRB,Zhang14_PRL,Sun14_PRB,Lahnsteiner_2022}. By projecting only onto the Raman active modes, we can isolate the corresponding part of the vibrational spectrum and in this way approximate Raman spectrum. Note however that the peak intensities do not correspond to those obtained from Raman and only the frequencies and lineshapes are relevant. 
Comparison to the total pDOS (i.e. containing both infrared and Raman active modes) is presented in the Supporting Information (see Fig.\ S5).

\begin{figure} 
    \centering
    \includegraphics[width=8.25cm]{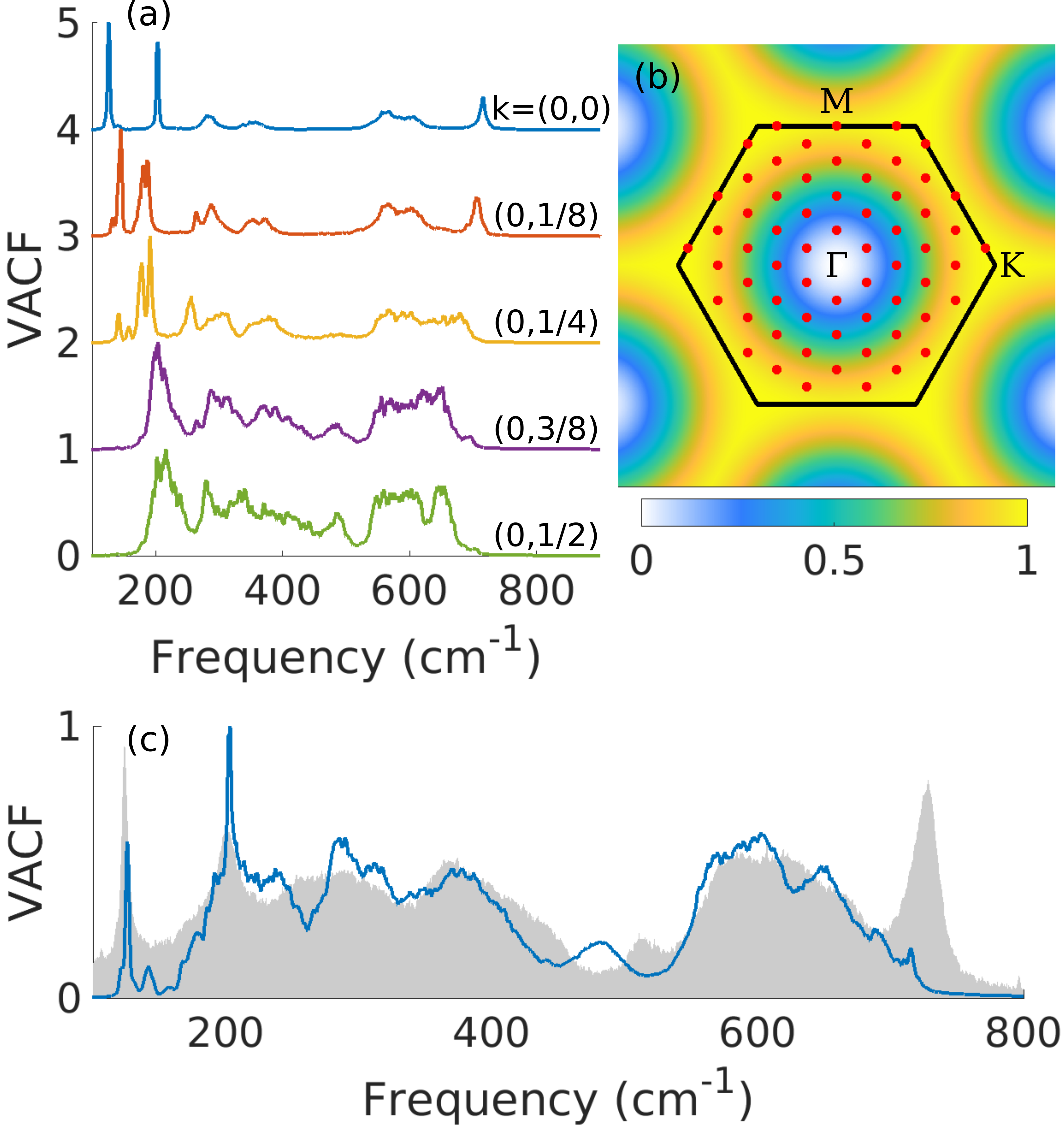}
    \caption{(a) Phonon density of states for wavevectors along the $\Gamma-\text{M}$ path. Labels on the right show the coordinates of the wavevector. (b) First Brillouin zone (black lines) with the 8x8 k-point mesh (red dots). The colormap represents the weights used in phonon-projected VACF. (c) Weighted phonon-projected VACF. Modes outside $\Gamma$ are weighted using the colormap of (b). The gray area shows the experimental Raman spectrum for comparison.}
    \label{fig:Fig5}
\end{figure}

We focus here on x=0.5 surfaces since we already determined this concentration as a good candidate to reproduce experimental results. 
Every point of the reciprocal space are not expected to contribute to the Raman spectra with the same weight. To find the weights, we study the pDOS for each wave vector along the $\Gamma-\text{M}$ path, which are shown in Fig.\ \ref{fig:Fig5}(a). pDOS at and close to $\Gamma$ mostly show low frequencies modes, coninciding with the Raman spectrum in Fig.\ \ref{fig:Fig4}. In experiment, these modes show narrow peaks which tends to indicate a small contribution from modes outside $\Gamma$. On the other hand, modes close to the edge of the first Brillouin zone show wide peaks around 300--400 and 600 cm$^{-1}$. Based on the experimental observations, we therefore choose larger weights for modes close to the BZ edges. The selected weights are represented in Fig.\ \ref{fig:Fig5}(b) and built from a sum of two Gaussians such that the weight approaches 0 at $\Gamma$-point and 1 at the BZ edge, except pDOS at $\Gamma$-point is also added with a weight of 1.

Resulting pDOS using these weights is represented in Fig.\ \ref{fig:Fig5}(c). When summing the weighted pDOS with pDOS at $\Gamma$, we find great agreement with experiment. The very wide peaks in the 300--400 and 600 cm$^{-1}$ regions are well reproduced. The peak at 200 cm$^{-1}$ also gets wider with the additional contributions from outside $\Gamma$, further improving the agreement with experiment. As previously discussed, pDOS does not reproduce well the intensities of peaks, which is especially true for the peak at 700 cm$^{-1}$. Also note that the spectrum now shows an additional peak around 500 cm$^{-1}$ which also appears in the experimental results, even though the agreement in frequency is not perfect. 
In Fig.\ \ref{fig:Fig4}(a), this peak shows up weakly in the OH-surfaces, but disappear in the mixed surfaces. However, since this mode is relatively flat [cf.\ the phonon dispersion in \ref{fig:Fig2}(c)], the inclusion of modes outside $\Gamma$-point could lead to increased intensity.
Thus, this peak seems to arise from the A$_\text{1g}$(OH) mode with its intensity enhanced by disorder, which we could confirm by inspecting the eigenmode-projected VACF in Fig.\ S5(b) in the Supporting Information. 

Similar pDOS for other concentrations is available in the Supporting Information (Fig.\ S6).
The results in Fig.\ S6 and those in Fig.\ \ref{fig:Fig4} suggest that the relative intensity of the two broad peaks at 300 \icm{} [E$_\text{g}$(OH)] and 400 \icm{} [E$_\text{g}$(O)] and of the two peaks that form the broad peak at 600 \icm{} [E$_\text{g}$(C)] could be used to probe the O/OH concentration, in addition to the shift of the sharp peaks.
In fact, our results are in agreement with all the changes with increasing O content (as verified using XPS) observed in the experiments by Lioi et al. \cite{Lioi_2019}: for the sharp peaks, there is downshift of 120 \icm{} [E$_\text{g}$(Ti)] peak, downshift of 200 \icm{} [A$_\text{1g}$(Ti)] peak, and upshift of 700 \icm{} [A$_\text{1g}$(C)] peak;
for the broad features, there is enhancement of the 400 \icm{} [E$_\text{g}$(O)] peak and enhancement of the lower frequency side of the 600 \icm{} peak. Only the 500 \icm{} [A$_\text{1g}$(OH)] does not show marked change.
The results in Figure S6 also illustrate that the surface concentration has a relatively minor impact on these peaks.
This finding, together with previous computational predictions of a limited range of accessible surface compositions under typical synthesis conditions \cite{Ibragimova19_ACSNano},
explains why the Raman spectra of \ce{Ti3C2T_{\rm{x}}} appears to depend weakly on the adopted synthesis procedure and on the sample morphology.


In conclusion, we have developed a computational approach to efficiently compute the Raman spectra of complex materials at finite temperature, which is based on a combination of machine-learning force fields and a method to efficiently evaluate the time-dependent susceptibility. Our approach can be applied to many other materials with high anharmonicity and pronounced temperature effects. Alternatively, it can be used for systems requiring large supercells when modeling, e.g.,alloys and/or defects. 

Here, we applied the approach to 2D titanium carbide MXene to study the effects of temperature and surface terminations on the Raman spectra. We find that the temperature plays a relatively small role and does not explain the wide peaks observed experimentally. 
On the other hand, mixed surfaces play an important role in the reproducing experimental Raman spectra. 
We identified three peaks that undergo straightforward evolution with the surface composition and thus could also be used to extract the composition from experimental spectra. Mixed surfaces lead to better agreement in the peak position for some peaks and, importantly, they also reproduce the two broad peaks observed at 300 and 400 cm$^{-1}$. Additionally, the modes outside $\Gamma$-point are found to play an important role, activated by the disorder such as defects.
In particular, modes close to the Brillouin zone edge are found to contribute significantly to the widening of peaks around 300--400 and 600 cm$^{-1}$, as well as the appearance of an additional peak around 500 cm$^{-1}$. Only by including the effect of heterogeneous surfaces and contribution from modes outside $\Gamma$-point, a good agreement with the experimentally measured spectra can be achieved.

\begin{acknowledgement}
We are grateful to the Academy of Finland for support under Academy Research Fellow funding No. 311058 and Academy Postdoc funding No. 330214. We also thank CSC-IT Center for Science Ltd. for generous grants of computer time.
\end{acknowledgement}

\begin{suppinfo}
Synthesis and Raman measurement of \ce{Ti3C2T_{\rm{x}}}; Resonant Raman effect; Comparison between distributions; Lorentzian fits of the peaks; Effect of multilayer and surface water; and Phonon density of states.
\end{suppinfo}

\bibliography{raman}

\providecommand{\latin}[1]{#1}
\makeatletter
\providecommand{\doi}
  {\begingroup\let\do\@makeother\dospecials
  \catcode`\{=1 \catcode`\}=2 \doi@aux}
\providecommand{\doi@aux}[1]{\endgroup\texttt{#1}}
\makeatother
\providecommand*\mcitethebibliography{\thebibliography}
\csname @ifundefined\endcsname{endmcitethebibliography}
  {\let\endmcitethebibliography\endthebibliography}{}
\begin{mcitethebibliography}{62}
\providecommand*\natexlab[1]{#1}
\providecommand*\mciteSetBstSublistMode[1]{}
\providecommand*\mciteSetBstMaxWidthForm[2]{}
\providecommand*\mciteBstWouldAddEndPuncttrue
  {\def\EndOfBibitem{\unskip.}}
\providecommand*\mciteBstWouldAddEndPunctfalse
  {\let\EndOfBibitem\relax}
\providecommand*\mciteSetBstMidEndSepPunct[3]{}
\providecommand*\mciteSetBstSublistLabelBeginEnd[3]{}
\providecommand*\EndOfBibitem{}
\mciteSetBstSublistMode{f}
\mciteSetBstMaxWidthForm{subitem}{(\alph{mcitesubitemcount})}
\mciteSetBstSublistLabelBeginEnd
  {\mcitemaxwidthsubitemform\space}
  {\relax}
  {\relax}

\bibitem[Naguib \latin{et~al.}(2021)Naguib, Barsoum, and Gogotsi]{Naguib_2021}
Naguib,~M.; Barsoum,~M.~W.; Gogotsi,~Y. Ten Years of Progress in the Synthesis
  and Development of MXenes. \emph{Advanced Materials} \textbf{2021},
  \emph{33}, 2103393\relax
\mciteBstWouldAddEndPuncttrue
\mciteSetBstMidEndSepPunct{\mcitedefaultmidpunct}
{\mcitedefaultendpunct}{\mcitedefaultseppunct}\relax
\EndOfBibitem
\bibitem[Babak \latin{et~al.}(2017)Babak, K., and Yuri]{Anasori_2017}
Babak,~A.; K.,~L.~M.; Yuri,~G. 2D metal carbides and nitrides (MXenes) for
  energy storage. \emph{Nature Reviews Materials} \textbf{2017}, 16098\relax
\mciteBstWouldAddEndPuncttrue
\mciteSetBstMidEndSepPunct{\mcitedefaultmidpunct}
{\mcitedefaultendpunct}{\mcitedefaultseppunct}\relax
\EndOfBibitem
\bibitem[Lee and Kim(2019)Lee, and Kim]{Lee_2019}
Lee,~E.; Kim,~D.-J. Review{\textemdash} Recent Exploration of Two-Dimensional
  {MXenes} for Gas Sensing: From a Theoretical to an Experimental View.
  \emph{Journal of The Electrochemical Society} \textbf{2019}, \emph{167},
  037515\relax
\mciteBstWouldAddEndPuncttrue
\mciteSetBstMidEndSepPunct{\mcitedefaultmidpunct}
{\mcitedefaultendpunct}{\mcitedefaultseppunct}\relax
\EndOfBibitem
\bibitem[Iqbal \latin{et~al.}(2020)Iqbal, Sambyal, and Koo]{Iqbal_2020}
Iqbal,~A.; Sambyal,~P.; Koo,~C.~M. 2D MXenes for Electromagnetic Shielding: A
  Review. \emph{Advanced Functional Materials} \textbf{2020}, \emph{30},
  2000883\relax
\mciteBstWouldAddEndPuncttrue
\mciteSetBstMidEndSepPunct{\mcitedefaultmidpunct}
{\mcitedefaultendpunct}{\mcitedefaultseppunct}\relax
\EndOfBibitem
\bibitem[Naguib \latin{et~al.}(2011)Naguib, Kurtoglu, Presser, Lu, Niu, Heon,
  Hultman, Gogotsi, and Barsoum]{Naguib_2011}
Naguib,~M.; Kurtoglu,~M.; Presser,~V.; Lu,~J.; Niu,~J.; Heon,~M.; Hultman,~L.;
  Gogotsi,~Y.; Barsoum,~M.~W. Two-Dimensional Nanocrystals Produced by
  Exfoliation of Ti3AlC2. \emph{Advanced Materials} \textbf{2011}, \emph{23},
  4248--4253\relax
\mciteBstWouldAddEndPuncttrue
\mciteSetBstMidEndSepPunct{\mcitedefaultmidpunct}
{\mcitedefaultendpunct}{\mcitedefaultseppunct}\relax
\EndOfBibitem
\bibitem[Mashtalir \latin{et~al.}(2013)Mashtalir, Naguib, Dyatkin, Gogotsi, and
  Barsoum]{Mashtalir13_MCP}
Mashtalir,~O.; Naguib,~M.; Dyatkin,~B.; Gogotsi,~Y.; Barsoum,~M.~W. Kinetics of
  aluminum extraction from Ti3AlC2 in hydrofluoric acid. \emph{Materials
  Chemistry and Physics} \textbf{2013}, \emph{139}, 147 -- 152\relax
\mciteBstWouldAddEndPuncttrue
\mciteSetBstMidEndSepPunct{\mcitedefaultmidpunct}
{\mcitedefaultendpunct}{\mcitedefaultseppunct}\relax
\EndOfBibitem
\bibitem[Shi \latin{et~al.}(2014)Shi, Beidaghi, Naguib, Mashtalir, Gogotsi, and
  Billinge]{Shi14_PRL}
Shi,~C.; Beidaghi,~M.; Naguib,~M.; Mashtalir,~O.; Gogotsi,~Y.; Billinge,~S.
  J.~L. Structure of Nanocrystalline ${\mathrm{Ti}}_{3}{\mathrm{C}}_{2}$ MXene
  Using Atomic Pair Distribution Function. \emph{Phys. Rev. Lett.}
  \textbf{2014}, \emph{112}, 125501\relax
\mciteBstWouldAddEndPuncttrue
\mciteSetBstMidEndSepPunct{\mcitedefaultmidpunct}
{\mcitedefaultendpunct}{\mcitedefaultseppunct}\relax
\EndOfBibitem
\bibitem[Wang \latin{et~al.}(2016)Wang, Naguib, Page, Wesolowski, and
  Gogotsi]{Wang16_CM}
Wang,~H.-W.; Naguib,~M.; Page,~K.; Wesolowski,~D.~J.; Gogotsi,~Y. Resolving the
  Structure of Ti3C2Tx MXenes through Multilevel Structural Modeling of the
  Atomic Pair Distribution Function. \emph{Chemistry of Materials}
  \textbf{2016}, \emph{28}, 349--359\relax
\mciteBstWouldAddEndPuncttrue
\mciteSetBstMidEndSepPunct{\mcitedefaultmidpunct}
{\mcitedefaultendpunct}{\mcitedefaultseppunct}\relax
\EndOfBibitem
\bibitem[Ibragimova \latin{et~al.}(2019)Ibragimova, Puska, and
  Komsa]{Ibragimova19_ACSNano}
Ibragimova,~R.; Puska,~M.~J.; Komsa,~H.-P. pH-Dependent Distribution of
  Functional Groups on Titanium-Based MXenes. \emph{ACS Nano} \textbf{2019},
  \emph{13}, 9171--9181\relax
\mciteBstWouldAddEndPuncttrue
\mciteSetBstMidEndSepPunct{\mcitedefaultmidpunct}
{\mcitedefaultendpunct}{\mcitedefaultseppunct}\relax
\EndOfBibitem
\bibitem[Ibragimova \latin{et~al.}(2021)Ibragimova, Erhart, Rinke, and
  Komsa]{Ibragimova21_JPCL}
Ibragimova,~R.; Erhart,~P.; Rinke,~P.; Komsa,~H.-P. Surface Functionalization
  of 2D MXenes: Trends in Distribution, Composition, and Electronic Properties.
  \emph{The Journal of Physical Chemistry Letters} \textbf{2021}, \emph{12},
  2377--2384\relax
\mciteBstWouldAddEndPuncttrue
\mciteSetBstMidEndSepPunct{\mcitedefaultmidpunct}
{\mcitedefaultendpunct}{\mcitedefaultseppunct}\relax
\EndOfBibitem
\bibitem[Sarycheva and Gogotsi(2020)Sarycheva, and Gogotsi]{Sarycheva_2020}
Sarycheva,~A.; Gogotsi,~Y. Raman Spectroscopy Analysis of the Structure and
  Surface Chemistry of Ti3C2Tx MXene. \emph{Chemistry of Materials}
  \textbf{2020}, \emph{32}, 3480--3488\relax
\mciteBstWouldAddEndPuncttrue
\mciteSetBstMidEndSepPunct{\mcitedefaultmidpunct}
{\mcitedefaultendpunct}{\mcitedefaultseppunct}\relax
\EndOfBibitem
\bibitem[Lioi \latin{et~al.}(2019)Lioi, Neher, Heckler, Back, Mehmood, Nepal,
  Pachter, Vaia, and Kennedy]{Lioi_2019}
Lioi,~D.~B.; Neher,~G.; Heckler,~J.~E.; Back,~T.; Mehmood,~F.; Nepal,~D.;
  Pachter,~R.; Vaia,~R.; Kennedy,~W.~J. Electron-Withdrawing Effect of Native
  Terminal Groups on the Lattice Structure of Ti3C2Tx MXenes Studied by
  Resonance Raman Scattering: Implications for Embedding MXenes in Electronic
  Composites. \emph{ACS Applied Nano Materials} \textbf{2019}, \emph{2},
  6087--6091\relax
\mciteBstWouldAddEndPuncttrue
\mciteSetBstMidEndSepPunct{\mcitedefaultmidpunct}
{\mcitedefaultendpunct}{\mcitedefaultseppunct}\relax
\EndOfBibitem
\bibitem[Sarycheva \latin{et~al.}(2022)Sarycheva, Shanmugasundaram, Krayev, and
  Gogotsi]{Sarycheva22_ACSNano}
Sarycheva,~A.; Shanmugasundaram,~M.; Krayev,~A.; Gogotsi,~Y. Tip-Enhanced Raman
  Scattering Imaging of Single- to Few-Layer Ti3C2Tx MXene. \emph{ACS Nano}
  \textbf{2022}, \emph{16}, 6858--6865, PMID: 35404582\relax
\mciteBstWouldAddEndPuncttrue
\mciteSetBstMidEndSepPunct{\mcitedefaultmidpunct}
{\mcitedefaultendpunct}{\mcitedefaultseppunct}\relax
\EndOfBibitem
\bibitem[Hu \latin{et~al.}(2015)Hu, Wang, Zhang, Li, Hu, and Wang]{Hu_2015}
Hu,~T.; Wang,~J.; Zhang,~H.; Li,~Z.; Hu,~M.; Wang,~X. Vibrational properties of
  Ti3C2 and Ti3C2T2 (T = O{,} F{,} OH) monosheets by first-principles
  calculations: a comparative study. \emph{Phys. Chem. Chem. Phys.}
  \textbf{2015}, \emph{17}, 9997--10003\relax
\mciteBstWouldAddEndPuncttrue
\mciteSetBstMidEndSepPunct{\mcitedefaultmidpunct}
{\mcitedefaultendpunct}{\mcitedefaultseppunct}\relax
\EndOfBibitem
\bibitem[Hu \latin{et~al.}(2018)Hu, Hu, Gao, Li, and Wang]{Hu_2018}
Hu,~T.; Hu,~M.; Gao,~B.; Li,~W.; Wang,~X. Screening Surface Structure of MXenes
  by High-Throughput Computation and Vibrational Spectroscopic Confirmation.
  \emph{The Journal of Physical Chemistry C} \textbf{2018}, \emph{122},
  18501--18509\relax
\mciteBstWouldAddEndPuncttrue
\mciteSetBstMidEndSepPunct{\mcitedefaultmidpunct}
{\mcitedefaultendpunct}{\mcitedefaultseppunct}\relax
\EndOfBibitem
\bibitem[Lindsay \latin{et~al.}(2013)Lindsay, Broido, and
  Reinecke]{Lindsay2013}
Lindsay,~L.; Broido,~D.~A.; Reinecke,~T.~L. Ab initio thermal transport in
  compound semiconductors. \emph{Phys. Rev. B} \textbf{2013}, \emph{87},
  165201\relax
\mciteBstWouldAddEndPuncttrue
\mciteSetBstMidEndSepPunct{\mcitedefaultmidpunct}
{\mcitedefaultendpunct}{\mcitedefaultseppunct}\relax
\EndOfBibitem
\bibitem[Thomas \latin{et~al.}(2013)Thomas, Brehm, Fligg, Vöhringer, and
  Kirchner]{Thomas13_PCCP}
Thomas,~M.; Brehm,~M.; Fligg,~R.; Vöhringer,~P.; Kirchner,~B. Computing
  vibrational spectra from ab initio molecular dynamics. \emph{Phys. Chem.
  Chem. Phys.} \textbf{2013}, \emph{15}, 6608--6622\relax
\mciteBstWouldAddEndPuncttrue
\mciteSetBstMidEndSepPunct{\mcitedefaultmidpunct}
{\mcitedefaultendpunct}{\mcitedefaultseppunct}\relax
\EndOfBibitem
\bibitem[Putrino and Parrinello(2002)Putrino, and Parrinello]{Putrino02_PRL}
Putrino,~A.; Parrinello,~M. Anharmonic Raman Spectra in High-Pressure Ice from
  Ab Initio Simulations. \emph{Phys. Rev. Lett.} \textbf{2002}, \emph{88},
  176401\relax
\mciteBstWouldAddEndPuncttrue
\mciteSetBstMidEndSepPunct{\mcitedefaultmidpunct}
{\mcitedefaultendpunct}{\mcitedefaultseppunct}\relax
\EndOfBibitem
\bibitem[Wan \latin{et~al.}(2013)Wan, Spanu, Galli, and Gygi]{Wan13_JCTC}
Wan,~Q.; Spanu,~L.; Galli,~G.~A.; Gygi,~F. Raman Spectra of Liquid Water from
  Ab Initio Molecular Dynamics: Vibrational Signatures of Charge Fluctuations
  in the Hydrogen Bond Network. \emph{Journal of Chemical Theory and
  Computation} \textbf{2013}, \emph{9}, 4124--4130\relax
\mciteBstWouldAddEndPuncttrue
\mciteSetBstMidEndSepPunct{\mcitedefaultmidpunct}
{\mcitedefaultendpunct}{\mcitedefaultseppunct}\relax
\EndOfBibitem
\bibitem[Luber \latin{et~al.}(2014)Luber, Iannuzzi, and Hutter]{Luber14_JCP}
Luber,~S.; Iannuzzi,~M.; Hutter,~J. Raman spectra from ab initio molecular
  dynamics and its application to liquid S-methyloxirane. \emph{The Journal of
  Chemical Physics} \textbf{2014}, \emph{141}, 094503\relax
\mciteBstWouldAddEndPuncttrue
\mciteSetBstMidEndSepPunct{\mcitedefaultmidpunct}
{\mcitedefaultendpunct}{\mcitedefaultseppunct}\relax
\EndOfBibitem
\bibitem[Kocer \latin{et~al.}(2022)Kocer, Ko, and Behler]{Kocer_2022}
Kocer,~E.; Ko,~T.~W.; Behler,~J. Neural Network Potentials: A Concise Overview
  of Methods. \emph{Annual Review of Physical Chemistry} \textbf{2022},
  \emph{73}, 163--186\relax
\mciteBstWouldAddEndPuncttrue
\mciteSetBstMidEndSepPunct{\mcitedefaultmidpunct}
{\mcitedefaultendpunct}{\mcitedefaultseppunct}\relax
\EndOfBibitem
\bibitem[Behler and Parrinello(2007)Behler, and Parrinello]{Behler_2007}
Behler,~J.; Parrinello,~M. Generalized Neural-Network Representation of
  High-Dimensional Potential-Energy Surfaces. \emph{Phys. Rev. Lett.}
  \textbf{2007}, \emph{98}, 146401\relax
\mciteBstWouldAddEndPuncttrue
\mciteSetBstMidEndSepPunct{\mcitedefaultmidpunct}
{\mcitedefaultendpunct}{\mcitedefaultseppunct}\relax
\EndOfBibitem
\bibitem[Bart\'ok \latin{et~al.}(2010)Bart\'ok, Payne, Kondor, and
  Cs\'anyi]{Bartok_2010}
Bart\'ok,~A.~P.; Payne,~M.~C.; Kondor,~R.; Cs\'anyi,~G. Gaussian Approximation
  Potentials: The Accuracy of Quantum Mechanics, without the Electrons.
  \emph{Phys. Rev. Lett.} \textbf{2010}, \emph{104}, 136403\relax
\mciteBstWouldAddEndPuncttrue
\mciteSetBstMidEndSepPunct{\mcitedefaultmidpunct}
{\mcitedefaultendpunct}{\mcitedefaultseppunct}\relax
\EndOfBibitem
\bibitem[Fan \latin{et~al.}(2021)Fan, Zeng, Zhang, Wang, Song, Dong, Chen, and
  Ala-Nissila]{Fan_2021}
Fan,~Z.; Zeng,~Z.; Zhang,~C.; Wang,~Y.; Song,~K.; Dong,~H.; Chen,~Y.;
  Ala-Nissila,~T. Neuroevolution machine learning potentials: Combining high
  accuracy and low cost in atomistic simulations and application to heat
  transport. \emph{Phys. Rev. B} \textbf{2021}, \emph{104}, 104309\relax
\mciteBstWouldAddEndPuncttrue
\mciteSetBstMidEndSepPunct{\mcitedefaultmidpunct}
{\mcitedefaultendpunct}{\mcitedefaultseppunct}\relax
\EndOfBibitem
\bibitem[Jinnouchi \latin{et~al.}(2019)Jinnouchi, Lahnsteiner, Karsai, Kresse,
  and Bokdam]{Jinnouchi_2019_1}
Jinnouchi,~R.; Lahnsteiner,~J.; Karsai,~F.; Kresse,~G.; Bokdam,~M. Phase
  Transitions of Hybrid Perovskites Simulated by Machine-Learning Force Fields
  Trained on the Fly with Bayesian Inference. \emph{Phys. Rev. Lett.}
  \textbf{2019}, \emph{122}, 225701\relax
\mciteBstWouldAddEndPuncttrue
\mciteSetBstMidEndSepPunct{\mcitedefaultmidpunct}
{\mcitedefaultendpunct}{\mcitedefaultseppunct}\relax
\EndOfBibitem
\bibitem[Jinnouchi \latin{et~al.}(2019)Jinnouchi, Karsai, and
  Kresse]{Jinnouchi_2019_2}
Jinnouchi,~R.; Karsai,~F.; Kresse,~G. On-the-fly machine learning force field
  generation: Application to melting points. \emph{Phys. Rev. B} \textbf{2019},
  \emph{100}, 014105\relax
\mciteBstWouldAddEndPuncttrue
\mciteSetBstMidEndSepPunct{\mcitedefaultmidpunct}
{\mcitedefaultendpunct}{\mcitedefaultseppunct}\relax
\EndOfBibitem
\bibitem[Hashemi \latin{et~al.}(2019)Hashemi, Krasheninnikov, Puska, and
  Komsa]{Hashemi19_PRM}
Hashemi,~A.; Krasheninnikov,~A.~V.; Puska,~M.; Komsa,~H.-P. Efficient method
  for calculating Raman spectra of solids with impurities and alloys and its
  application to two-dimensional transition metal dichalcogenides. \emph{Phys.
  Rev. Materials} \textbf{2019}, \emph{3}, 023806\relax
\mciteBstWouldAddEndPuncttrue
\mciteSetBstMidEndSepPunct{\mcitedefaultmidpunct}
{\mcitedefaultendpunct}{\mcitedefaultseppunct}\relax
\EndOfBibitem
\bibitem[Oliver \latin{et~al.}(2020)Oliver, Fox, Hashemi, Singh, Cavalero, Yee,
  Snyder, Jaramillo, Komsa, and Vora]{Oliver20_JMCC}
Oliver,~S.~M.; Fox,~J.~J.; Hashemi,~A.; Singh,~A.; Cavalero,~R.~L.; Yee,~S.;
  Snyder,~D.~W.; Jaramillo,~R.; Komsa,~H.-P.; Vora,~P.~M. Phonons and excitons
  in ZrSe2–ZrS2 alloys. \emph{J. Mater. Chem. C} \textbf{2020}, \emph{8},
  5732--5743\relax
\mciteBstWouldAddEndPuncttrue
\mciteSetBstMidEndSepPunct{\mcitedefaultmidpunct}
{\mcitedefaultendpunct}{\mcitedefaultseppunct}\relax
\EndOfBibitem
\bibitem[Kou \latin{et~al.}(2020)Kou, Hashemi, Puska, Krasheninnikov, and
  Komsa]{Kou_2020}
Kou,~Z.; Hashemi,~A.; Puska,~M.; Krasheninnikov,~A.~V.; Komsa,~H.-P. Efficient
  method for calculating Raman spectra of solids with impurities and alloys and
  its application to two-dimensional transition metal dichalcogenides.
  \emph{npj Comput Mater} \textbf{2020}, \emph{6}, 59\relax
\mciteBstWouldAddEndPuncttrue
\mciteSetBstMidEndSepPunct{\mcitedefaultmidpunct}
{\mcitedefaultendpunct}{\mcitedefaultseppunct}\relax
\EndOfBibitem
\bibitem[Sutter \latin{et~al.}(2021)Sutter, Komsa, Lu, Gruverman, and
  Sutter]{Sutter21_NT}
Sutter,~P.; Komsa,~H.; Lu,~H.; Gruverman,~A.; Sutter,~E. Few-layer tin sulfide
  (SnS): Controlled synthesis, thickness dependent vibrational properties, and
  ferroelectric ity. \emph{Nano Today} \textbf{2021}, \emph{37}, 101082\relax
\mciteBstWouldAddEndPuncttrue
\mciteSetBstMidEndSepPunct{\mcitedefaultmidpunct}
{\mcitedefaultendpunct}{\mcitedefaultseppunct}\relax
\EndOfBibitem
\bibitem[Medders and Paesani(2015)Medders, and Paesani]{Medders_2015}
Medders,~G.~R.; Paesani,~F. Infrared and Raman Spectroscopy of Liquid Water
  through “First-Principles” Many-Body Molecular Dynamics. \emph{Journal of
  Chemical Theory and Computation} \textbf{2015}, \emph{11}, 1145--1154\relax
\mciteBstWouldAddEndPuncttrue
\mciteSetBstMidEndSepPunct{\mcitedefaultmidpunct}
{\mcitedefaultendpunct}{\mcitedefaultseppunct}\relax
\EndOfBibitem
\bibitem[Placzek(1934)]{placzek1934}
Placzek,~G. \emph{Rayleigh-streuung und Raman-effekt}; Akademische
  Verlagsgesellschaft, 1934; Vol.~2\relax
\mciteBstWouldAddEndPuncttrue
\mciteSetBstMidEndSepPunct{\mcitedefaultmidpunct}
{\mcitedefaultendpunct}{\mcitedefaultseppunct}\relax
\EndOfBibitem
\bibitem[Long(2002)]{Long2002-qo}
Long,~D.~A. \emph{The Raman effect}; John Wiley \& Sons: Chichester, England,
  2002\relax
\mciteBstWouldAddEndPuncttrue
\mciteSetBstMidEndSepPunct{\mcitedefaultmidpunct}
{\mcitedefaultendpunct}{\mcitedefaultseppunct}\relax
\EndOfBibitem
\bibitem[Kresse and Furthmüller(1996)Kresse, and Furthmüller]{Kresse_1996_1}
Kresse,~G.; Furthmüller,~J. Efficiency of ab-initio total energy calculations
  for metals and semiconductors using a plane-wave basis set.
  \emph{Computational Materials Science} \textbf{1996}, \emph{6}, 15--50\relax
\mciteBstWouldAddEndPuncttrue
\mciteSetBstMidEndSepPunct{\mcitedefaultmidpunct}
{\mcitedefaultendpunct}{\mcitedefaultseppunct}\relax
\EndOfBibitem
\bibitem[Kresse and Furthm\"uller(1996)Kresse, and
  Furthm\"uller]{Kresse_1996_2}
Kresse,~G.; Furthm\"uller,~J. Efficient iterative schemes for ab initio
  total-energy calculations using a plane-wave basis set. \emph{Phys. Rev. B}
  \textbf{1996}, \emph{54}, 11169--11186\relax
\mciteBstWouldAddEndPuncttrue
\mciteSetBstMidEndSepPunct{\mcitedefaultmidpunct}
{\mcitedefaultendpunct}{\mcitedefaultseppunct}\relax
\EndOfBibitem
\bibitem[Perdew \latin{et~al.}(2008)Perdew, Ruzsinszky, Csonka, Vydrov,
  Scuseria, Constantin, Zhou, and Burke]{Perdew_2008}
Perdew,~J.~P.; Ruzsinszky,~A.; Csonka,~G.~I.; Vydrov,~O.~A.; Scuseria,~G.~E.;
  Constantin,~L.~A.; Zhou,~X.; Burke,~K. Restoring the Density-Gradient
  Expansion for Exchange in Solids and Surfaces. \emph{Phys. Rev. Lett.}
  \textbf{2008}, \emph{100}, 136406\relax
\mciteBstWouldAddEndPuncttrue
\mciteSetBstMidEndSepPunct{\mcitedefaultmidpunct}
{\mcitedefaultendpunct}{\mcitedefaultseppunct}\relax
\EndOfBibitem
\bibitem[Togo and Tanaka(2015)Togo, and Tanaka]{Togo_2015}
Togo,~A.; Tanaka,~I. First principles phonon calculations in materials science.
  \emph{Scr. Mater.} \textbf{2015}, \emph{108}, 1--5\relax
\mciteBstWouldAddEndPuncttrue
\mciteSetBstMidEndSepPunct{\mcitedefaultmidpunct}
{\mcitedefaultendpunct}{\mcitedefaultseppunct}\relax
\EndOfBibitem
\bibitem[Walter and Moseler(2020)Walter, and Moseler]{Walter20_JCTC}
Walter,~M.; Moseler,~M. Ab Initio Wavelength-Dependent Raman Spectra: Placzek
  Approximation and Beyond. \emph{Journal of Chemical Theory and Computation}
  \textbf{2020}, \emph{16}, 576--586, PMID: 31815473\relax
\mciteBstWouldAddEndPuncttrue
\mciteSetBstMidEndSepPunct{\mcitedefaultmidpunct}
{\mcitedefaultendpunct}{\mcitedefaultseppunct}\relax
\EndOfBibitem
\bibitem[Gajdo\ifmmode~\check{s}\else \v{s}\fi{}
  \latin{et~al.}(2006)Gajdo\ifmmode~\check{s}\else \v{s}\fi{}, Hummer, Kresse,
  Furthm\"uller, and Bechstedt]{Gajdos_2006}
Gajdo\ifmmode~\check{s}\else \v{s}\fi{},~M.; Hummer,~K.; Kresse,~G.;
  Furthm\"uller,~J.; Bechstedt,~F. Linear optical properties in the
  projector-augmented wave methodology. \emph{Phys. Rev. B} \textbf{2006},
  \emph{73}, 045112\relax
\mciteBstWouldAddEndPuncttrue
\mciteSetBstMidEndSepPunct{\mcitedefaultmidpunct}
{\mcitedefaultendpunct}{\mcitedefaultseppunct}\relax
\EndOfBibitem
\bibitem[Liu \latin{et~al.}(2021)Liu, Verdi, Karsai, and Kresse]{Liu_2021}
Liu,~P.; Verdi,~C.; Karsai,~F.; Kresse,~G.
  $\ensuremath{\alpha}\text{\ensuremath{-}}\ensuremath{\beta}$ phase transition
  of zirconium predicted by on-the-fly machine-learned force field. \emph{Phys.
  Rev. Materials} \textbf{2021}, \emph{5}, 053804\relax
\mciteBstWouldAddEndPuncttrue
\mciteSetBstMidEndSepPunct{\mcitedefaultmidpunct}
{\mcitedefaultendpunct}{\mcitedefaultseppunct}\relax
\EndOfBibitem
\bibitem[Bokdam \latin{et~al.}(2021)Bokdam, Lahnsteiner, and
  Sarma]{Bokdam_2021}
Bokdam,~M.; Lahnsteiner,~J.; Sarma,~D.~D. Exploring Librational Pathways with
  on-the-Fly Machine-Learning Force Fields: Methylammonium Molecules in MAPbX3
  (X = I, Br, Cl) Perovskites. \emph{The Journal of Physical Chemistry C}
  \textbf{2021}, \emph{125}, 21077--21086\relax
\mciteBstWouldAddEndPuncttrue
\mciteSetBstMidEndSepPunct{\mcitedefaultmidpunct}
{\mcitedefaultendpunct}{\mcitedefaultseppunct}\relax
\EndOfBibitem
\bibitem[Lahnsteiner and Bokdam(2022)Lahnsteiner, and Bokdam]{Lahnsteiner_2022}
Lahnsteiner,~J.; Bokdam,~M. Anharmonic lattice dynamics in large thermodynamic
  ensembles with machine-learning force fields:
  $\mathrm{Cs}\mathrm{Pb}{\mathrm{Br}}_{3}$, a phonon liquid with Cs rattlers.
  \emph{Phys. Rev. B} \textbf{2022}, \emph{105}, 024302\relax
\mciteBstWouldAddEndPuncttrue
\mciteSetBstMidEndSepPunct{\mcitedefaultmidpunct}
{\mcitedefaultendpunct}{\mcitedefaultseppunct}\relax
\EndOfBibitem
\bibitem[Xie and Kent(2013)Xie, and Kent]{Xie13_PRB}
Xie,~Y.; Kent,~P. R.~C. Hybrid density functional study of structural and
  electronic properties of functionalized Ti${}_{n+1}{X}_{n}$ ($X=\mathrm{C}$,
  N) monolayers. \emph{Phys. Rev. B} \textbf{2013}, \emph{87}, 235441\relax
\mciteBstWouldAddEndPuncttrue
\mciteSetBstMidEndSepPunct{\mcitedefaultmidpunct}
{\mcitedefaultendpunct}{\mcitedefaultseppunct}\relax
\EndOfBibitem
\bibitem[Khazaei \latin{et~al.}(2017)Khazaei, Ranjbar, Arai, Sasaki, and
  Yunoki]{Khazaei17_JMCC}
Khazaei,~M.; Ranjbar,~A.; Arai,~M.; Sasaki,~T.; Yunoki,~S. Electronic
  properties and applications of MXenes: a theoretical review. \emph{J. Mater.
  Chem. C} \textbf{2017}, \emph{5}, 2488--2503\relax
\mciteBstWouldAddEndPuncttrue
\mciteSetBstMidEndSepPunct{\mcitedefaultmidpunct}
{\mcitedefaultendpunct}{\mcitedefaultseppunct}\relax
\EndOfBibitem
\bibitem[Hu \latin{et~al.}(2016)Hu, Hu, Li, Zhang, Zhang, Wang, and
  Wang]{Hu16_PCCP}
Hu,~T.; Hu,~M.; Li,~Z.; Zhang,~H.; Zhang,~C.; Wang,~J.; Wang,~X. Interlayer
  coupling in two-dimensional titanium carbide MXenes. \emph{Phys. Chem. Chem.
  Phys.} \textbf{2016}, \emph{18}, 20256--20260\relax
\mciteBstWouldAddEndPuncttrue
\mciteSetBstMidEndSepPunct{\mcitedefaultmidpunct}
{\mcitedefaultendpunct}{\mcitedefaultseppunct}\relax
\EndOfBibitem
\bibitem[Nosé(1984)]{Nose_1984}
Nosé,~S. A unified formulation of the constant temperature molecular dynamics
  methods. \emph{The Journal of Chemical Physics} \textbf{1984}, \emph{81},
  511--519\relax
\mciteBstWouldAddEndPuncttrue
\mciteSetBstMidEndSepPunct{\mcitedefaultmidpunct}
{\mcitedefaultendpunct}{\mcitedefaultseppunct}\relax
\EndOfBibitem
\bibitem[Shuichi(1991)]{Nose_1991}
Shuichi,~N. {Constant Temperature Molecular Dynamics Methods}. \emph{Progress
  of Theoretical Physics Supplement} \textbf{1991}, \emph{103}, 1--46\relax
\mciteBstWouldAddEndPuncttrue
\mciteSetBstMidEndSepPunct{\mcitedefaultmidpunct}
{\mcitedefaultendpunct}{\mcitedefaultseppunct}\relax
\EndOfBibitem
\bibitem[Zhang \latin{et~al.}(2020)Zhang, Hu, Wang, and Zhou]{Zhang20_JMST}
Zhang,~H.; Hu,~T.; Wang,~X.; Zhou,~Y. Structural defects in MAX phases and
  their derivative MXenes: A look forward. \emph{Journal of Materials Science
  \& Technology} \textbf{2020}, \emph{38}, 205 -- 220\relax
\mciteBstWouldAddEndPuncttrue
\mciteSetBstMidEndSepPunct{\mcitedefaultmidpunct}
{\mcitedefaultendpunct}{\mcitedefaultseppunct}\relax
\EndOfBibitem
\bibitem[He \latin{et~al.}(2019)He, Wan, Zhao, Guo, Jiang, and Zheng]{He19_CTC}
He,~R.; Wan,~Y.; Zhao,~P.; Guo,~P.; Jiang,~Z.; Zheng,~J. First-principles
  investigation of native point defects in two-dimensional Ti3C2.
  \emph{Computational and Theoretical Chemistry} \textbf{2019}, \emph{1150}, 26
  -- 39\relax
\mciteBstWouldAddEndPuncttrue
\mciteSetBstMidEndSepPunct{\mcitedefaultmidpunct}
{\mcitedefaultendpunct}{\mcitedefaultseppunct}\relax
\EndOfBibitem
\bibitem[Ibragimova \latin{et~al.}(2022)Ibragimova, Rinke, and
  Komsa]{Ibragimova22_CM}
Ibragimova,~R.; Rinke,~P.; Komsa,~H.-P. Native Vacancy Defects in MXenes at
  Etching Conditions. \emph{Chemistry of Materials} \textbf{2022}, \emph{34},
  2896--2906\relax
\mciteBstWouldAddEndPuncttrue
\mciteSetBstMidEndSepPunct{\mcitedefaultmidpunct}
{\mcitedefaultendpunct}{\mcitedefaultseppunct}\relax
\EndOfBibitem
\bibitem[Benitez \latin{et~al.}(2016)Benitez, Kan, Gao, O'Neal, Proust, and
  Radovic]{Benitez16_AM}
Benitez,~R.; Kan,~W.~H.; Gao,~H.; O'Neal,~M.; Proust,~G.; Radovic,~M. Room
  temperature stress-strain hysteresis in Ti2AlC revisited. \emph{Acta
  Materialia} \textbf{2016}, \emph{105}, 294--305\relax
\mciteBstWouldAddEndPuncttrue
\mciteSetBstMidEndSepPunct{\mcitedefaultmidpunct}
{\mcitedefaultendpunct}{\mcitedefaultseppunct}\relax
\EndOfBibitem
\bibitem[Karlsson \latin{et~al.}(2015)Karlsson, Birch, Halim, Barsoum, and
  Persson]{Karlsson15_NL}
Karlsson,~L.~H.; Birch,~J.; Halim,~J.; Barsoum,~M.~W.; Persson,~P. O.~{\AA}.
  Atomically Resolved Structural and Chemical Investigation of Single MXene
  Sheets. \emph{Nano Letters} \textbf{2015}, \emph{15}, 4955--4960\relax
\mciteBstWouldAddEndPuncttrue
\mciteSetBstMidEndSepPunct{\mcitedefaultmidpunct}
{\mcitedefaultendpunct}{\mcitedefaultseppunct}\relax
\EndOfBibitem
\bibitem[Sang \latin{et~al.}(2016)Sang, Xie, Lin, Alhabeb, Van~Aken, Gogotsi,
  Kent, Xiao, and Unocic]{Sang16_ACSNano}
Sang,~X.; Xie,~Y.; Lin,~M.-W.; Alhabeb,~M.; Van~Aken,~K.~L.; Gogotsi,~Y.;
  Kent,~P. R.~C.; Xiao,~K.; Unocic,~R.~R. Atomic Defects in Monolayer Titanium
  Carbide (Ti3C2Tx) MXene. \emph{ACS Nano} \textbf{2016}, \emph{10},
  9193--9200, PMID: 27598326\relax
\mciteBstWouldAddEndPuncttrue
\mciteSetBstMidEndSepPunct{\mcitedefaultmidpunct}
{\mcitedefaultendpunct}{\mcitedefaultseppunct}\relax
\EndOfBibitem
\bibitem[Mathis \latin{et~al.}(2021)Mathis, Maleski, Goad, Sarycheva, Anayee,
  Foucher, Hantanasirisakul, Shuck, Stach, and Gogotsi]{Mathis21_ACSNano}
Mathis,~T.~S.; Maleski,~K.; Goad,~A.; Sarycheva,~A.; Anayee,~M.;
  Foucher,~A.~C.; Hantanasirisakul,~K.; Shuck,~C.~E.; Stach,~E.~A.; Gogotsi,~Y.
  Modified MAX Phase Synthesis for Environmentally Stable and Highly Conductive
  Ti3C2 MXene. \emph{ACS Nano} \textbf{2021}, \emph{15}, 6420--6429, PMID:
  33848136\relax
\mciteBstWouldAddEndPuncttrue
\mciteSetBstMidEndSepPunct{\mcitedefaultmidpunct}
{\mcitedefaultendpunct}{\mcitedefaultseppunct}\relax
\EndOfBibitem
\bibitem[Tian \latin{et~al.}(2022)Tian, Ju, Luo, Bin, Lou, and Que]{Tian22_CEJ}
Tian,~Y.; Ju,~M.; Luo,~Y.; Bin,~X.; Lou,~X.; Que,~W. In situ oxygen doped
  Ti3C2Tx MXene flexible film as supercapacitor electrode. \emph{Chemical
  Engineering Journal} \textbf{2022}, \emph{446}, 137451\relax
\mciteBstWouldAddEndPuncttrue
\mciteSetBstMidEndSepPunct{\mcitedefaultmidpunct}
{\mcitedefaultendpunct}{\mcitedefaultseppunct}\relax
\EndOfBibitem
\bibitem[Cardona and Brodsky(1982)Cardona, and Brodsky]{cardona1982}
Cardona,~M.; Brodsky,~M. \emph{Light Scattering in Solids}; Light Scattering in
  Solids nid. 2; Springer-Verlag, 1982\relax
\mciteBstWouldAddEndPuncttrue
\mciteSetBstMidEndSepPunct{\mcitedefaultmidpunct}
{\mcitedefaultendpunct}{\mcitedefaultseppunct}\relax
\EndOfBibitem
\bibitem[Ushioda(1974)]{Ushioda_1974}
Ushioda,~S. Defect-activated first order Raman scattering in boron implanted
  GaAs. \emph{Solid State Communications} \textbf{1974}, \emph{15},
  149--153\relax
\mciteBstWouldAddEndPuncttrue
\mciteSetBstMidEndSepPunct{\mcitedefaultmidpunct}
{\mcitedefaultendpunct}{\mcitedefaultseppunct}\relax
\EndOfBibitem
\bibitem[Eckmann \latin{et~al.}(2013)Eckmann, Felten, Verzhbitskiy, Davey, and
  Casiraghi]{Eckmann_2013}
Eckmann,~A.; Felten,~A.; Verzhbitskiy,~I.; Davey,~R.; Casiraghi,~C. Raman study
  on defective graphene: Effect of the excitation energy, type, and amount of
  defects. \emph{Phys. Rev. B} \textbf{2013}, \emph{88}, 035426\relax
\mciteBstWouldAddEndPuncttrue
\mciteSetBstMidEndSepPunct{\mcitedefaultmidpunct}
{\mcitedefaultendpunct}{\mcitedefaultseppunct}\relax
\EndOfBibitem
\bibitem[Wang \latin{et~al.}(1990)Wang, Chan, and Ho]{Wang90_PRB}
Wang,~C.~Z.; Chan,~C.~T.; Ho,~K.~M. Tight-binding molecular-dynamics study of
  phonon anharmonic effects in silicon and diamond. \emph{Phys. Rev. B}
  \textbf{1990}, \emph{42}, 11276--11283\relax
\mciteBstWouldAddEndPuncttrue
\mciteSetBstMidEndSepPunct{\mcitedefaultmidpunct}
{\mcitedefaultendpunct}{\mcitedefaultseppunct}\relax
\EndOfBibitem
\bibitem[Zhang \latin{et~al.}(2014)Zhang, Sun, and Wentzcovitch]{Zhang14_PRL}
Zhang,~D.-B.; Sun,~T.; Wentzcovitch,~R.~M. Phonon Quasiparticles and Anharmonic
  Free Energy in Complex Systems. \emph{Phys. Rev. Lett.} \textbf{2014},
  \emph{112}, 058501\relax
\mciteBstWouldAddEndPuncttrue
\mciteSetBstMidEndSepPunct{\mcitedefaultmidpunct}
{\mcitedefaultendpunct}{\mcitedefaultseppunct}\relax
\EndOfBibitem
\bibitem[Sun \latin{et~al.}(2014)Sun, Zhang, and Wentzcovitch]{Sun14_PRB}
Sun,~T.; Zhang,~D.-B.; Wentzcovitch,~R.~M. Dynamic stabilization of cubic
  $\mathrm{Ca}\mathrm{Si}{\mathrm{O}}_{3}$ perovskite at high temperatures and
  pressures from ab initio molecular dynamics. \emph{Phys. Rev. B}
  \textbf{2014}, \emph{89}, 094109\relax
\mciteBstWouldAddEndPuncttrue
\mciteSetBstMidEndSepPunct{\mcitedefaultmidpunct}
{\mcitedefaultendpunct}{\mcitedefaultseppunct}\relax
\EndOfBibitem
\end{mcitethebibliography}

\newpage

\

\newpage

\includepdf[pages=-]{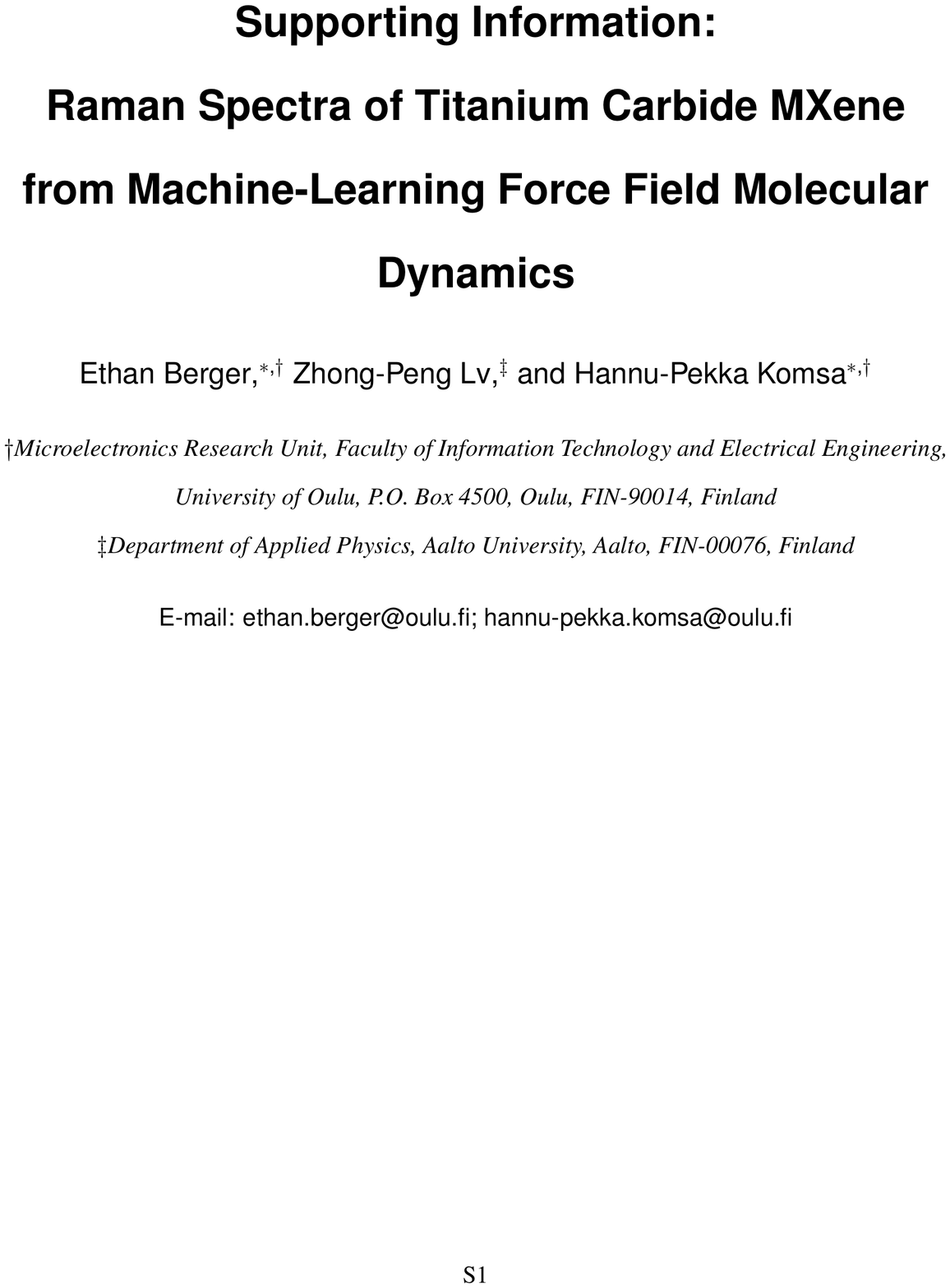}

\end{document}